\documentclass[prl,twocolumn,nofootinbib,aps,tightenlines,preprintnumbers,notitlepage,longbibliography,superscriptaddress]{revtex4-1}
\usepackage{graphicx}
\usepackage{amsmath}
\usepackage{amsfonts}
\usepackage[colorlinks=true,citecolor=blue,urlcolor=cyan,linkcolor=black]{hyperref}

\newcommand{\nhat}{\hat{ \mathbf{n}}}

\usepackage{amsmath}
\usepackage{amsfonts}

\usepackage{tabularx}
\newcolumntype{C}{>{\centering\arraybackslash}X}
\newcolumntype{R}{>{\raggedleft\arraybackslash}X}

\usepackage[usenames, dvipsnames]{color}
\usepackage[normalem]{ulem}
\usepackage{xcolor}

\newcommand{\dd}{{\rm d}}

\def\nhat{\hat{\mathbf{n}}}

\newcommand{\be}{\begin{eqnarray}}

\newcommand{\ee}{\end{eqnarray}}
\newcommand{\wj}[6]{\left(
                           \begin{array}{ccc}
        \! #1\! & #2\!  & #3\!\!  \\
        \! #4\! & #5\!  & #6\!\!
                           \end{array}
                   \right)}

\newcommand{\imperial}{Department of Physics, Imperial College London, Blackett Laboratory, Prince Consort Road, London SW7 2AZ, UK}

\newcommand{\perimeter}{Perimeter Institute for Theoretical Physics, 31 Caroline St N, Waterloo, ON N2L 2Y5, Canada}

\newcommand{\york}{Department of Physics and Astronomy, York University, Toronto, ON M3J 1P3, Canada}

\graphicspath{ {plots/} }

\begin{document}

\title{Reconstructing large scales at cosmic dawn}

\author{Selim~C.~Hotinli}
\affiliation{\imperial}

\author{Matthew~C.~Johnson}
\affiliation{\perimeter}
\affiliation{\york}

\begin{abstract}
The cosmic microwave background (CMB) serves as a backlight to large-scale structure during the epoch of reionization, where Thomson scattering gives rise to temperature anisotropies on small angular scales from the kinetic Sunyaev Zel'dovich (kSZ) effect. In this paper, we demonstrate that the technique of kSZ tomography (velocity reconstruction), based on cross correlations between CMB temperature and 21cm surveys,  can significantly improve constraints on models of inhomogeneous reionization and provide information about large-scale modes that are poorly characterized by 21cm measurements themselves due to foreground contamination.
\end{abstract}

\maketitle

\textbf{\textit{Introduction}} -- The epoch of reionization (EoR), corresponding to the time when the first stars formed (cosmic dawn) and ionized the majority of neutral Hydrogen in the Universe, is among the least understood parts of our cosmic history. Excitingly, surveys of the redshifted 21cm hydrogen-line such as HERA~\citep{DeBoer:2016tnn} and SKA~\citep{Bacon:2018dui}, cosmic microwave background (CMB) experiments such as the Simons Observatory (SO)~\citep{Ade:2018sbj} and CMB-S4~\citep{Abazajian:2016yjj}, as well as infrared, X-ray, and line intensity mapping missions promise to deliver an abundance of information on the EoR in the coming decade. There is much to learn about astrophysics from EoR measurements. Additionally, because it is a high redshift probe, the EoR in principle carries information about cosmology. In this paper, we propose a new method for extracting both astrophysical and cosmological information from the EoR by combining data from 21cm and CMB surveys using the technique of kinetic Sunyaev Zel'dovich (kSZ) tomography~\cite{Zhang:2010fc,Terrana:2016xvc,Deutsch:2017ybc,Smith:2018bpn}. 

A few hundred million years after the Big Bang, the Universe went through a phase of patchy reionization, where local bubbles of ionized gas formed around the first stars initiate, and eventually complete, the transition to a fully-ionized baryonic component of the Universe. As patchy reionization unfolds, CMB photons Thompson scatter from free electrons inside the bubbles, giving rise to temperature anisotropies proportional to the local density and the CMB dipole observed in the rest frame of the electrons~\cite{Valageas:2000ke,Zhang:2003nr,McQuinn:2005ce,Iliev:2006un}. This is the kSZ effect~\citep{Zeldovich:1969en,Sunyaev:1970er,Sunyaev:1980vz,1986ApJ...306L..51O}, referred to here as `reionization kSZ' to distinguish the present context from lower redshift contributions to the kSZ temperature anisotropies. Reionization kSZ makes an important blackbody contribution to the observed CMB temperature anisotropies on small angular scales, and with detection imminent in the next generation CMB experiments, there have been significant recent efforts to model reionization kSZ with increasing accuracy (see e.g.~\cite{Alvarez:2015xzu,Park:2017amo}) and devise statistics to extract information about reionization (see e.g. \citep{Alvarez:2020gvl,Smith:2016lnt,Ferraro:2018izc}). 

Given a tracer of the ionized bubbles formed during patchy reionization, we show that it is possible to extract information about the remote dipole field (the projected CMB dipole observed along our past light cone; see e.g. Refs.~\cite{Zhang:2015uta,Deutsch:2017ybc}) from the measured reionization kSZ temperature anisotropies. We develop a quadratic estimator~\cite{Deutsch:2017ybc,Smith:2018bpn} for the dipole field based on the correlations between the redshifted 21cm hydrogen-line, which is anti-correlated with ionized regions~\cite{Cooray:2004ei,Salvaterra:2005js}, and CMB temperature anisotropies on small angular scales. Because reionization occurs at relatively high redshfits ($z\sim 7.5$, corresponding to a radial comoving distance of $\sim 9$ Gpc), the remote dipole field contains information about inhomogeneities on very large physical scales, making it a useful cosmological probe. Indeed, direct measurements of 21cm on large scales can improve constraints on e.g. the optical depth~\cite{Liu:2015txa}, $\Lambda$CDM~\cite{McQuinn:2005hk}, neutrino masses~\cite{Pritchard:2008wy}, isocurvature~\cite{Gordon_2009}, or the running of the spectral index~\cite{Adshead:2010mc}. However, using 21cm on large scales to infer the underlying density field will be difficult due to foreground contamination~\citep{Pober:2012zz}, making the information in the reconstructed dipole field highly valuable and entirely complementary to direct 21cm observations. 

Cross correlations between 21 cm and kSZ also provide constraints on models of reionization. Previous literature has considered constraints deriving from the kSZ-21cm cross power~\cite{Alvarez:2005sa,Adshead:2007ij,Giannantonio:2007za,2010MNRAS.402.2279J,2010MNRAS.402.2617T} as well as higher order statistics such as the kSZ-kSZ-21cm bispectrum~\cite{Ma:2017gey,LaPlante:2020nxx}. In this paper, we propose that correlations between the remote dipole field and 21cm observations can provide tight constraints on models of reionization. This is equivalent to the 21cm-21cm-kSZ bispectrum, in complete analogy with recent results on the late-time kSZ effect~\cite{Smith:2018bpn}, which in the context of reionization was shown in Ref.~\cite{Cooray:2004ei} to contain more information than the kSZ-21cm cross power. Another probe of reionization comes from small-angular scale measurements of CMB polarization, which can be used to reconstruct the inhomogeneous optical depth during reionization~\cite{Dvorkin:2008tf}. We show that including correlations between the reconstructed optical depth, 21cm, and the remote dipole field improves constraints due to complementary parameter degeneracies.

\vspace{0.1cm}

\noindent \textbf{\textit{Observables}} -- Including fluctuations in the ionization fraction along a line of sight $\nhat$ at a radial comoving distance $\chi$, the optical depth is defined as:
\be\label{eq:optdepth}
\tau(\chi\nhat) = - \sigma_T \int d\chi \ a(\chi) \bar{n}_b (\chi) \left[ \bar{x}_e(\chi)+\delta x_e(\chi\nhat) \right] \,. 
\ee
In terms of redshift $z$, the mean ionization fraction during reionization is 
\be\label{eq:mean_reio}
\overline{x}_e(z)=\frac{1}{2}\left[1-\tanh{\left(\frac{y(z)-y_{\rm re}}{\Delta_y}\right)}\right]\,,
\ee
where $\Delta_y$ and $y_{\rm re}$ are model parameters and $y(z)=(1+z)^{3/2}$. We trade $y_{\rm err}$ for the mean optical depth $\tau = - \sigma_T \int d\chi \ a(\chi) \bar{n}_b (\chi) \bar{x}_e(\chi)$. We model the inhomogeneities in the ionization fraction following Refs.~\cite{Furlanetto:2004nh,Zahn:2006sg,Dvorkin:2008tf} as due to bubbles whose radius $R$ follows a log-normal distribution with mean size $\bar{R}$ and width $\sigma_{\ln R}$
\be
P(R)\!=\!R^{-1}(2\pi^2\sigma_{\ln R}^2)^{-0.5}\!\exp\{\!-\!\ln[(R/\bar{R})^2]/(2\sigma_{\ln R}^2)\}.\ \ \
\ee
Finally, we assume the number density of bubbles fluctuates as a biased tracer of the large-scale structure with a bubble bias, $b$. We set the fiducial values of our reionization model as $\{\tau,\Delta_y,b,\bar{R},\sigma_{\ln R}\}
=\{0.06,7.0,6.0,5\ {\rm Mpc},0.7\}$. Note that we translate the constraints on $\Delta_y$ into another common parametrisation, $\Delta z=z|_{\bar{x}_e=0.75}-z|_{\bar{x}_e=0.25}$, with fiducial value $\Delta z_0=1.66$ for our parameter choices. 

\noindent{{\textit{Hydrogen power-spectra}}\label{sec:hydrogen_power}} --  Surveys of 21cm hydrogen-line measure the brightness temperature. We model the amplitude of the brightness temperature fluctuation, $\delta T_{21}(\chi\nhat) \equiv \overline{T}_{21}(\chi)-T_{21}(\chi\nhat)$, as proportional to the neutral hydrogen density fluctuation, 
$\delta T_{21}(\chi\nhat)=F^{1/2}(\chi)\delta_{\rm H}(\chi\nhat)\,,$
where in terms of redshift
\be
F^{1/2}(z)=0.023{\rm K} \frac{1-Y_p}{0.75}\frac{\Omega_b h^2}{0.02}\left(\frac{1+z}{10}\frac{0.15}{\Omega_mh^2}\right)\,,
\ee
and $Y_p$ is the primordial helium fraction. Note we ignored modelling complicating factors such as the baryonic feedback, spatial fluctuations in spin temperature and the effect of redshift-space distortions, which are beyond the scope of our paper. We account for these astrophysical uncertainties by marginalizing over the amplitude of the 21cm hydrogen spectra when forecasting.

The neutral hydrogen fluctuations can be separated into contributions from the ionization fraction and the gas density function, $\delta(\chi\nhat)$, as~e.g.~in~\citep{Zaldarriaga:2003du,Furlanetto:2004nh,Wang:2005my},
\be
\delta_{\rm H}(\chi\nhat)=[1-\bar{x}_e(\chi\nhat)]\delta(\chi\nhat)-\delta x_{e}(\chi\nhat)\,.
\ee
Ionized hydrogen fluctuations, or equivalently the free electron, satisfy
$\delta_{ e}(\chi\nhat)= \bar{x}_e\delta(\chi\nhat)+\delta x_e(\chi\nhat)$.

Following~\citep{Zaldarriaga:2003du,Furlanetto:2004nh,Wang:2005my}, we work in the context of the halo model~\citep{Cooray:2002dia} and express power spectra in terms of correlations between (2-bubble) and within (1-bubble) reionization bubbles. For the forecasts presented below, we need the power spectra for Hydrogen $P_{\rm HH} (k,z) = \langle \delta_{\rm H} (k,z)^2 \rangle$, electrons $P_{ee} (k,z) = \langle \delta_e (k,z)^2 \rangle$, and cross-powers, $P_{e\rm H}$ and $P_{\delta\rm H}$. The large-scale 2-bubble hydrogen-intensity and electron power-spectra are: 
\be
P_{\rm HH}^{2b}=[(1-\bar{x}_e)(\ln(1-\bar{x}_e)b\langle W_R(k)\rangle +1)]^2P_{\delta\delta}(k)\,.\ \ \ \  
\ee
\be
P_{ee}^{2b}(z)=[(1-\bar{x}_e)\ln(1-\bar{x}_e) b\langle W_R(k)\rangle-\bar{x}_e]^2P_{\delta\delta}(k)\,.\ \ \ \
\ee
where $P_{\delta\delta}(k)$ is the linear matter power-spectrum, 
\be
\langle W_R(k) \rangle = \langle V_b \rangle^{-1}\int\dd R P(R) V_b(R)W_R(k)\,,
\ee
where $\langle V_b \rangle \equiv \int\dd R P(R) V_b(R)$, and
\be
W_R(k)\equiv {3}(kR)^{-1}\left[\sin(kR)-kR\cos(kR)\right]\,.
\ee
Note we omit showing redshift dependence of the ionization fraction and the power-spectra for notational brevity. Following~\cite{Wang:2005my}, the small-scale one-bubble power spectra are:
\be
P_{\rm HH}^{1b}=P_{ee}^{1b}\simeq\bar{x}_e(1-\bar{x}_e)[\langle V_b\rangle\langle W_R^2(k)\rangle+\tilde{P}(k)]\,,
\ee
where $\langle W_R^2(k)\rangle\equiv{\langle V_b\rangle^{-2}}\!\!\int\!\dd R V_b^2(R)P(R)W_R^2(k)$, and 
\be
\tilde{P}(k)\simeq{P(k)\langle V_b\rangle\langle \sigma^2_R\rangle}{[P(k)^2+(\langle V_b\rangle \langle \sigma_R^2\rangle)^2]^{-1/2}}\!,\ 
\ee
and $\langle \sigma_R^2 \rangle$ is the smoothed density variance averaged over the bubble radius distribution,
\be
\langle\sigma_R^2\rangle={\langle V_b\rangle^{-2}}\!\int\!\dd R\, V_b^2(R)P(R)\,\sigma_R^2\,.
\ee
Finally, the cross-correlations between the matter density, the neutral hydrogen and the free electron fluctuations can be found as $P_{e\rm H}^{2b}(k)  =P_{\rm H\delta}(k)-P_{\rm HH}^{2b}(k)\,,$ where
\be
P_{\rm H\delta}(k)=(1-\bar{x}_e)[\ln(1-\bar{x}_e)b\langle W_R(k)\rangle-1] P_{\delta \delta}(k)\,,\ \ \ \ 
\ee
and $P_{e\rm H}^{1b}(k)=-P^{1b}_{ee}(k)$.

\begin{figure}[t!]
    \vspace{-0.25cm}
    \includegraphics[width=\columnwidth]{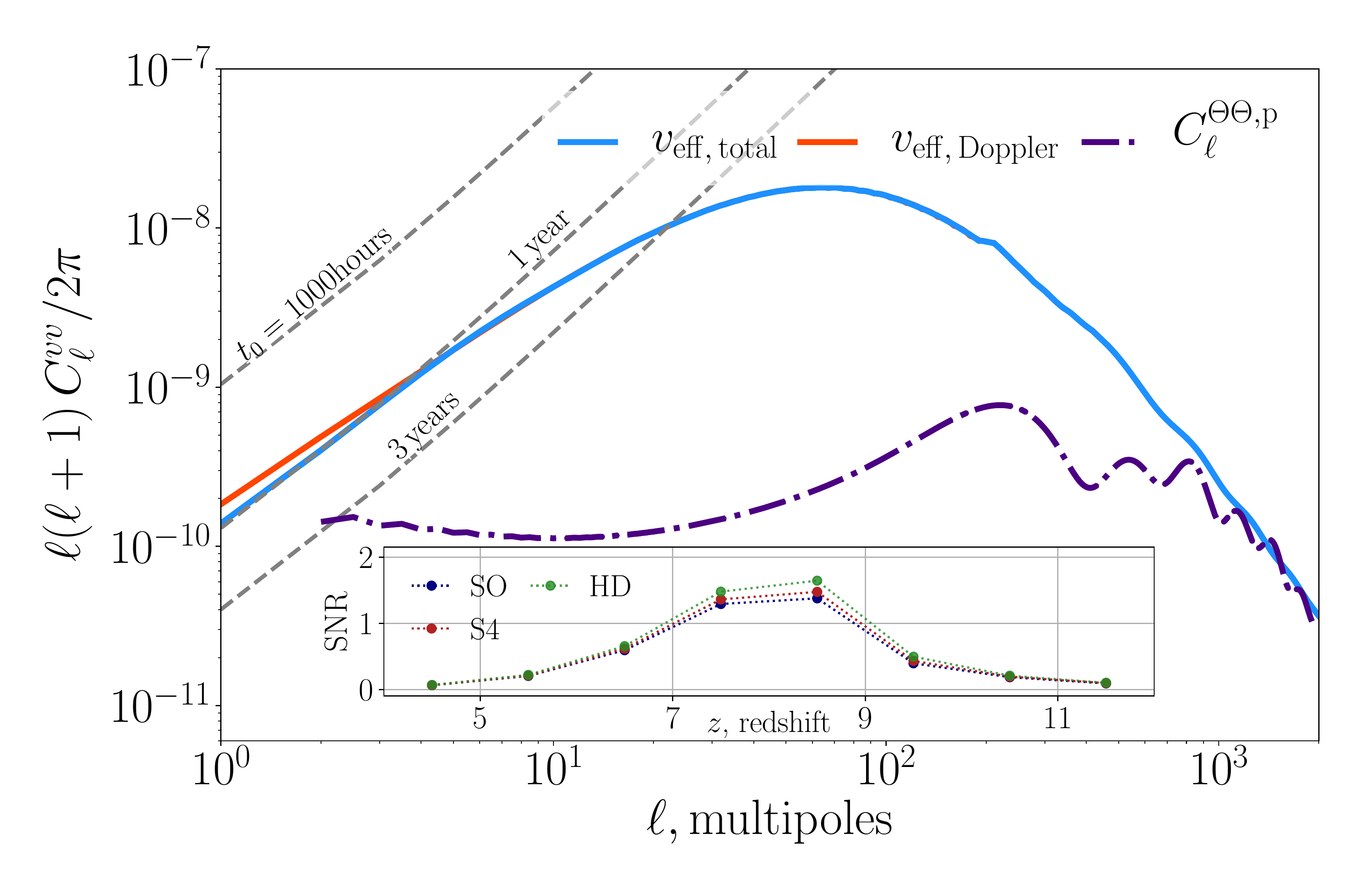}
    \vspace{-0.75cm}
    \caption{Dipole field in a redshift bin centred at $z=8.5$ and of size $\Delta z=1$ (blue, solid), shown with its Doppler component (orange, solid) and the primary CMB (purple, dot-dashed). Dashed gray lines show reconstruction noise forecasts from the combination of CMB S4 and SKA survey, latter for varying integration times $t_0=\{1000\,{\rm hours},\,1\,{\rm year},\,3\,{\rm years}\}$ from top to bottom. The inset plot shows the SNR forecasts per redshift bins of size $\Delta z=1$, for varying CMB experiments and 1000\,hours of measurement time for a SKA-like experiment.}
    \label{fig:recon_noise_limits}
\end{figure}

\noindent{\textit{{Temperature and Polarization Anisotropies}}\label{sec:The_kSZ_effect}} --  Temperature anisotropies in the $\nhat$ direction are sourced during reionization by the kSZ effect and the screening of the primary CMB temperature anisotropies $ \Theta_p (\nhat)$~\citep{2009PhRvD..79j7302D,Dvorkin:2008tf}
\begin{eqnarray}
\Theta_{\rm rei}(\nhat) &\!\simeq &\!\int_{\rm rei} \!\!\!\dd\chi \ \dot{\tau} (\chi\nhat) e^{-\tau(\chi)} \,\left[ v_{\rm eff}(\chi\nhat)\!-\!\Theta_p (\nhat) \right] \\
&\simeq& \sum_\alpha  \left[\bar{v}_{\rm eff}^\alpha (\nhat)\!-\!\Theta_p (\nhat) \right]\!\int_{\chi^\alpha_{\rm min}}^{\chi^\alpha_{\rm max}}\!\!\!\dd\chi \ \!\dot{\tau} (\chi\nhat) e^{-\tau(\chi)}\,\,\,\,\,\,\,\,\,\\
&\simeq&  \sum_\alpha  \left[\bar{v}_{\rm eff}^\alpha (\nhat)\!-\!\Theta_p (\nhat) \right] \Delta\tau^\alpha (\nhat)\,,
\end{eqnarray}
where the optical depth and its $\chi$ derivative, $\dot{\tau}$, are defined by Eq.~\eqref{eq:optdepth} and $v_{\rm eff}=3\int\dd^2\nhat_e\Theta_1(\chi \nhat,\nhat_e)\nhat\cdot\nhat_e/(4\pi)$ is the remote dipole field projected along the line of sight; note that we neglect the evolution of the primary CMB between reionization and the present day (e.g. due to the ISW effect). In the second line, we bin the contribution to the visibility, approximate the remote dipole by its bin average, and neglect terms beyond linear order in $\tau$. Note that while the dominant contribution to the remote dipole field is the Doppler effect due to local peculiar velocity, there is a significant primordial contribution from the Sachs Wolfe effect on the largest angular scales. We contrast the full and Doppler contribution to the remote dipole power spectrum in Fig.~\ref{fig:recon_noise_limits}; we also show the primary CMB temperature fluctuations for reference.

Analogously, the polarization anisotropies during reionization are sourced by Thomson, the polarized Sunyaev Zel'dovich (pSZ) effect, and screening of polarization anisotropies produced at recombination. The contributions due to inhomogeneous reionization are:
\begin{eqnarray}
\Theta^{\pm}_{\rm rei} (\nhat) &\simeq& \sum_\alpha  \left[ \bar{q}_{\rm eff}^{\pm; \alpha}(\nhat) - \Theta^{\pm}_{\rm rec} (\nhat) \right] \Delta \tau^\alpha (\nhat)
\end{eqnarray}
where $\Theta^{\pm} \equiv (Q \pm iU)$ and $\bar{q}_{\rm eff}^{\pm; \alpha}(\nhat)$ is the bin-averaged remote quadrupole field (the locally observed CMB quadrupole).

\noindent\textbf{\textit{Reconstruction}}
\label{sec:ksz_Estimator} --The temperature and polarization anisotropies sourced during patchy reionization from a redshift bin are products of the the anisotropic optical depth $\Delta \tau^\alpha (\nhat)$ and the difference of the dipole or quadrupole fields and primordial temperature or polarization anisotropies. Given the dipole/quadrupole fields and the temperature/polarization anisotropies, Ref.~\cite{Dvorkin:2008tf} constructed a quadratic estimator for the anisotropic optical depth $\Delta \tau (\nhat) \equiv \sum_\alpha \Delta \tau^\alpha (\nhat)$ (in fact, a weighted sum is reconstructed; we neglect this complication for the moment). Turning this around, Ref.~\cite{Meerburg:2017lfh} constructed a quadratic estimator for the polarization anisotropies given a tracer of the anisotropic optical depth. In this paper, we propose a new quadratic estimator for the remote dipole field using the 21cm line as a (redshift-dependent) tracer of the anisotropic optical depth. In analogy with the estimator for the late-time kSZ effect~\citep{Deutsch:2017ybc}, the estimator for the averaged remote dipole field over redshift bin $\alpha$ is:
\be
\label{eq:estimator1}
&&\widehat{v}_{{\rm eff}, \ell m}^{\alpha}\\ 
&&\!=\!b_v^\alpha {N^{vv}_{\alpha \ell}}\!\!\!\!\!\!\sum_{\ell_1m_1\ell_2m_2}\!\!\!\!\!\!\!(-1)^m
\Gamma_{\ell_1\ell_2\ell}^\alpha
\!\wj{\ell_1}{\ell_2}{\ell}{m_1}{m_2}{-m}\!\! 
\frac{a^{\Theta}_{\ell_1m_1} \delta^{\alpha}_{H,\ell_2m_2}}{C^{\Theta\Theta\rm,obs}_{\ell_1} C^{\rm HH,obs}_{\alpha \ell_2} }, \nonumber
\ee
where $b_v^\alpha$ is the ``optical depth bias'', which, in our study, is due to mismodelling $C^{\tau \rm H}_{\alpha,\ell_2}$ in the presence of foregrounds, baryonic feedback and other factors that bias the relation between the observed temperature brightness and the hydrogen density, $\delta T_{21}\rightarrow b_H F^{1/2}(z)\delta_{\rm H}(\chi\nhat)$, and the optical depth bias is a function of $b_{\rm H}$ whose functional form is shown in Eq.~(69) of Ref.~\citep{Smith:2018bpn}. The coefficient is 

\begin{equation}
\Gamma_{\ell_1\ell_2 \ell}^ \alpha = \sqrt{\frac{(2\ell_1+1)(2\ell_2+1)(2\ell+1)}{4\pi}} \wj{\ell_1}{\ell_2}{\ell}{0}{0}{0} \ C^{\tau \rm H}_{\alpha,\ell_2}\,,
\end{equation}
and the reconstruction noise (i.e. variance of the estimator) is defined by
\begin{equation}
\label{eq:estimator1noise}
\frac{1}{N^{vv}_{\alpha \ell}} = \frac{1}{(2\ell+1)} \sum_{\ell_1\ell_2}
\frac{\Gamma_{\ell_1\ell_2\ell}^\alpha \ \Gamma_{\ell_1\ell_2\ell}^\alpha}{C^{\Theta\Theta,\rm obs}_{\ell_1} C^{\rm HH,obs}_{\alpha \ell_2} }. 
\end{equation}
In these expressions, $C^{\Theta\Theta,\rm obs}_{\ell_1}$ is the measured CMB temperature power spectrum, $C^{\rm HH,obs}_{\alpha \ell_2}$ is the measured spectrum of the mean 21cm hydrogen fluctuations in each redshift bin, ${\delta}_{\rm H,\alpha}^{\,\mathrm{obs}}(\chi\nhat)=\int^{\chi_{\rm max}^\alpha}_{\chi_{\rm min}^\alpha} \dd \chi W_{\alpha}(\chi){\delta}_{\rm H}^{\,\mathrm{obs}}(\chi\nhat)$, where $W_\alpha(\chi)$ is a top-hat selection function for redshift bin $\alpha$, and $C^{\tau \rm H}_{\alpha,\ell_2}$ is the cross-power of the optical depth and brightness temperature counts in each bin. In principle, the reconstruction noise improves by probing increasing small angular scales with the CMB and brightness temperature, and is limited only by the vanishing $\tau$ and $H$ correlation on very small scales. In reality, the reconstruction noise is limited by the instrumental noise of the CMB experiment and thermal noise of the 21cm experiment, since this places an effective upper limit in $\ell$ on the sum in Eq.~\eqref{eq:estimator1noise}. Due to the contribution from screening, the reconstruction in each bin will be biased by $\Theta_p(\nhat)$; this can be subtracted using our knowledge of the well-measured primary CMB on large angular scales.

Below, we also make use of the reconstructed anisotropic optical depth from measurements of CMB polarization. The quadratic estimator for $\Delta \tau (\nhat)$ was given in Ref.~\cite{Dvorkin:2008tf}; the variance is:
\be
N_{\ell}^{\tau\tau}\!=\!\left[\frac{1}{2\ell+1}\sum\limits_{\ell_1\ell_2}\frac{|\Gamma_{\ell_1\ell_2\ell}^{EB}|^2}{(C_{\ell_1}^{EE}+N_{\ell_1}^{EE})(C_{\ell_2}^{BB}+N_{\ell_2}^{BB})}\right]^{-1}\!\!\!\!\!\!\! .\,\,\,\,\,\,\,\,
\ee
with 
\be
\Gamma_{\ell_1\ell_2\ell}^{EB}&=&\frac{C_{\ell_1}^{E_0E_1}}{2i}\sqrt{\frac{(2\ell_1+)(2\ell_2+1)(2\ell+1)}{4\pi}}\nonumber\\
&&\ \ \times\left[\wj{\ell_1}{\ell_2}{\ell}{-2}{2}{0}-\wj{\ell_1}{\ell_2}{\ell}{2}{-2}{0}\right]\,,
\ee
where $C_{\ell_1}^{E_0 E_1}$ is the cross power between the E-mode polarization anisotropies with ($E_1$) and without ($E_0$) the contributions from patchy reionization.

\noindent\textbf{\textit{Forecasts}\label{sec:forecasts}} -- We model the experimental noise from the 21cm brightness measurement as in~\citep{Ansari:2018ury}, with experimental specifications appropriate to the upcoming SKA survey. We use the lensed $C_\ell^{TT}$, include the kSZ effect~\citep{Shaw:2011sy,George:2014oba} and implement a realistic consideration of ILC-cleaned foregrounds with configurations anticipated for SO~\citep{Ade:2018sbj}, CMB-S4~\citep{Abazajian:2019eic} and a futuristic high-definition experiment, CMB-HD~\citep{Sehgal:2019ewc,private_mat}. We assume a joint $40$ percent sky coverage between the 21cm experiments and SO, CMB-S4 and CMB-HD, respectively. 

We find, for 8 redshift bins inside the range $z\in[4,12]$, and for a total integration time of 1000 hours for the 21cm experiment, the total detection SNR (assuming zero null-condition in the reconstructed dipole field) is $\{3.6,3.8,4.1\}$ for SO, CMB-S4 and CMB-HD, respectively. We show the detection SNR per redshift bin in the inset of Fig.~\ref{fig:recon_noise_limits}. We find for a full 3-year 21cm experiment, the SNR can reach $\{18.6,19.4,21.0\}$ for the same set of experiments. We find that in all cases the 21cm experimental noise dominates the kSZ reconstruction error. 

\begin{figure}[b!]
    \vspace{-0.25cm}
    \includegraphics[width=1\columnwidth]{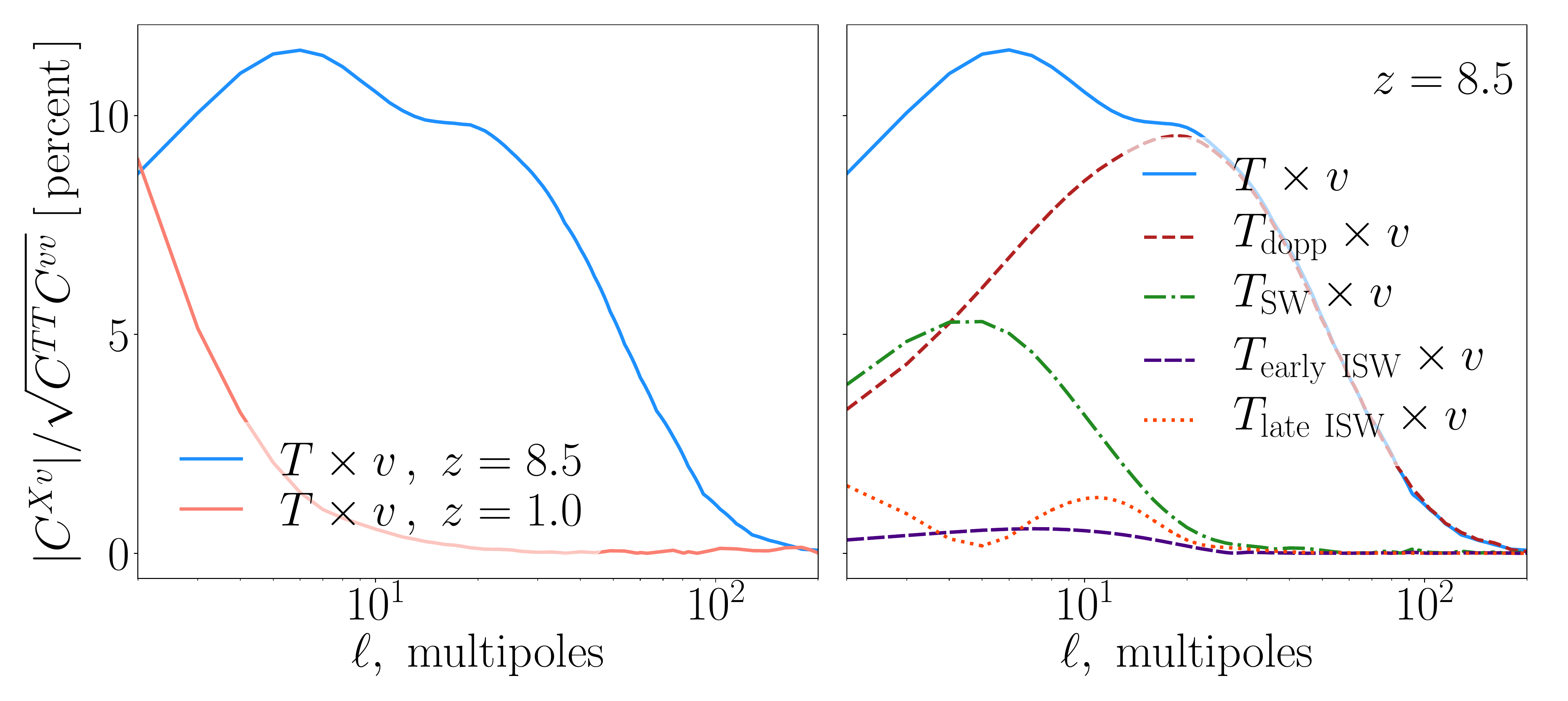}
    \vspace{-0.75cm}
    \caption{ (\textit{Left}) Correlation coefficient, $|C^{Xv}|/\!\sqrt{C^{TT}\!C^{vv}}$, between the CMB temperature and the remote dipole at two redshifts $z\in\{1.0,8.5\}$. For higher redshifts remote dipole probes CMB at higher accuracy and precision. (\textit{Right}) Correlation coefficient between varying contributions to the CMB temperature $X\in\{T_{\rm dopp},T_{\rm SW},T_{\rm ISW}\}$ and the remote dipole. The larger range of $\ell$, accessible to higher redshifts, allow probing different components of the CMB at different scales.}
    \label{fig:x_corr_per_ell}
\end{figure}

With an SKA-like experiment, it is possible to produce maps of the dipole field over a significant range of scales. This is illustrated by Fig.~\ref{fig:recon_noise_limits}, which compares the remote dipole power spectrum to the reconstruction noise in a redshift bin centred at $z=8.5$. The cosmological value of the dipole field reconstructed at lower redshift has been established in many recent studies~\cite{Hotinli:2019wdp,Sato-Polito:2020cil,Pan:2019dax,Contreras:2019bxy,Cayuso:2019hen,Munchmeyer:2018eey,Zhang:2015uta}. Maps of the remote dipole field at reionization probe larger scales, which we illustrate by computing the correlation coefficient between the dipole field at a variety of redshfits and the primary CMB, as shown in Fig.~\ref{fig:x_corr_per_ell}. During reionization, there is correlation at the $\sim 10 \%$ level with a variety of the contributions to the primary CMB over a significant range of multipoles. We can therefore conclude that the remote dipole field contains new information on scales comparable to the primary CMB, making it a promising observable to improve upon cosmic-variance limited constraints from the primary CMB alone. Furthermore, the correlation structure between the remote dipole field and the various components of the CMB (right panel of Fig.~\ref{fig:x_corr_per_ell}) can potentially be used to break degeneracies suffered by other probes of cosmology, and to improve constraints on parameters sensitive to large scale CMB and density fluctuations. We leave a detailed study to future work.

Turning to potential constraints on reionization, we construct the Fisher matrix incorporating various combinations of 21cm, the remote dipole, and the reconstructed optical depth
\be
\label{eq:fisherM}
F_{ab}\!=\!\!\!\sum_{\ell=\ell_{\rm min}}^{\ell_{{\rm max}}}\!\!f_{\rm sky}\frac{2\ell+1}{2}{\rm Tr}\left[\left(\partial_{a}\mathbf{C}_{\ell}\right)\mathbf{C}_{\ell}^{-1}\left(\partial_{b}\mathbf{C}_{\ell}\right)\mathbf{C}_{\ell}^{-1}\right],\ \ \ \ 
\ee
where indices $\{a,b\}$ represent reionization model parameters introduced earlier (we fix LCDM parameters to \textit{Planck} 2018 cosmology~\citep{Aghanim:2018eyx}). We assume the Fisher matrix for the optical dept reconstruction to be independent from the kSZ tomography, and set the covariance matrix, $\mathbf{C_\ell}$, for the latter to include auto-correlations of the neutral hydrogen and the remote dipole, as well as their correlations, between 8 redshift bins. We add an additional bias parameter for the temperature brightness at every redshift bin, effectively marginalising over both the amplitude of $\delta_{\rm H}(\chi\nhat)$ and the optical depth bias from the kSZ, which are excpected to be subject to a large model uncertainties. 

\begin{figure}[h!]
    \vspace{-0.25cm}
    \includegraphics[width=1\columnwidth]{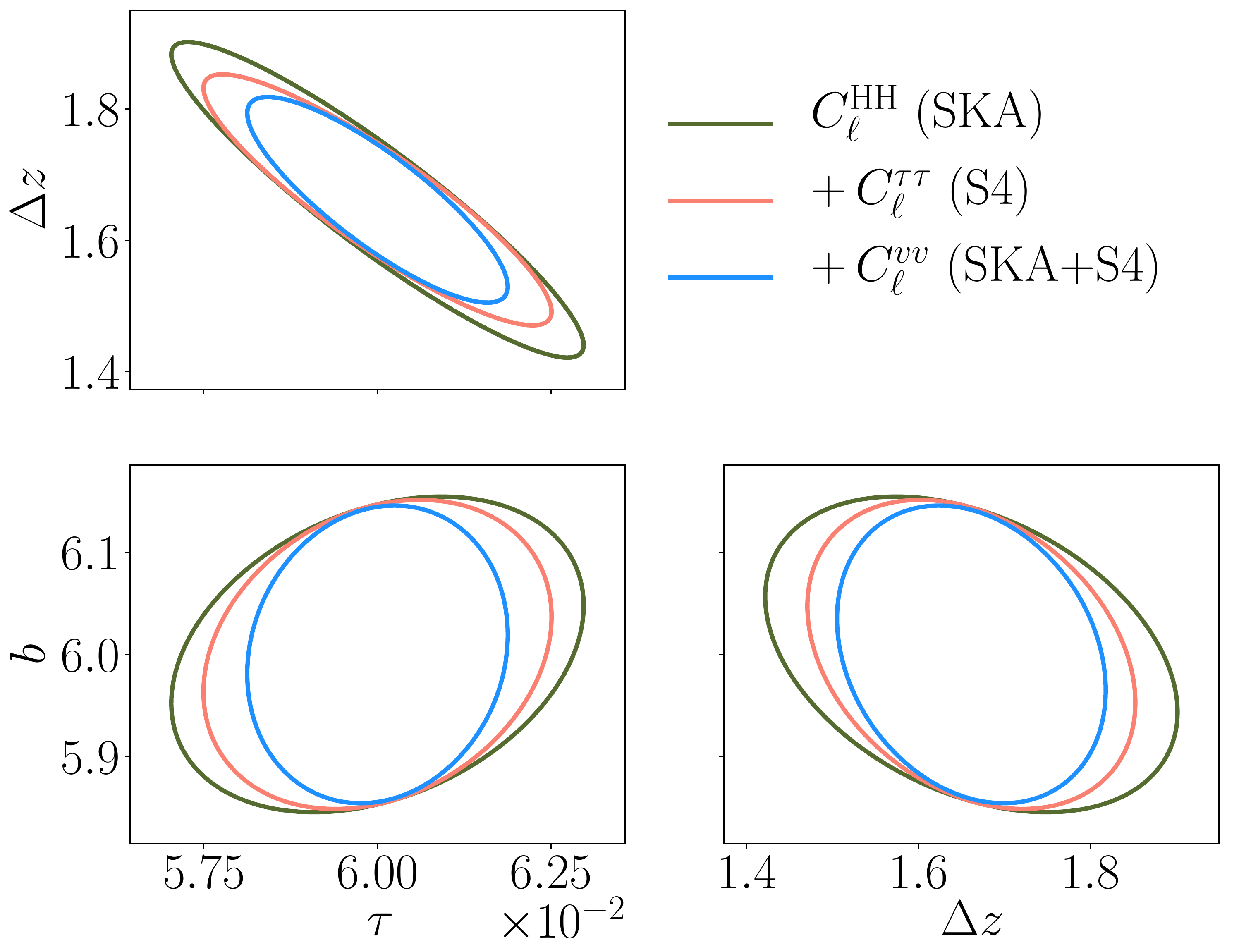}
    \vspace{-0.75cm}
    \caption{Fisher forecasts on reionization parameters for 68$\%$ confidence-limit from kSZ-reconstructed large-scale velocity fluctuations, reconstructed anisotropic $\tau$-field, and the neutral hydrogen fluctuations on large scales, $k_{\rm min}\chi<\ell<300$, where we apply a large-scale cut-off at $k_{\rm min}=0.01h/$Mpc at all redshift bins to approximate the anticipated foreground wedge (see~e.g.~\citep{Pober:2012zz}). For the velocity reconstruction, we use 8 redshift bins in the range $z\in[4,12]$. The 21cm experiment specifications are chosen similar to that of SKA, with $40\%$ mutual sky coverage with the CMB experiment. The SKA measurements assume 1000 hours of observation time.}
    \label{fig:Fisher_1}
\end{figure}

We show our results on Fig.~\ref{fig:Fisher_1} for the anticipated noise levels from an SKA-like experiment, with 1000 hours of integration time, together with the constraints from optical depth reconstruction from the CMB. We find, with a CMB-S4-like experiment, $\sigma(\tau)\!\lesssim\!4.0\!\times\!10^{-3}$ from measurement of $C_\ell^{\rm HH}$ alone, and $\sigma(\tau)\!\lesssim\!4.5\!\times\!10^{-3}$ from $C_\ell^{v\rm H}$ and $C_\ell^{vv}$ together. Combined constraints from the hydrogen and remote dipole fields satisfy $\sigma(\tau)\!\lesssim\!2.2\!\times10^{-3}$, comparable to the cosmic-variance limit that can be achieved from large-angle CMB polarization data. Fig.~\ref{fig:Fisher_1} suggests the potential improvement on the reionization parameters from the kSZ reconstruction is comparable to the $\tau$ reconstruction and the measurement of hydrogen density fluctuations alone, while better modelling the baryonic feedback mechanisms can potentially increase the constraining power of the hydrogen density beyond the velocities and the $\tau$-field. 

\noindent\textbf{\textit{Discussion  --}}\label{sec:discussion} In this work, we discussed the scientific value of the cross-correlation between the kSZ effect, which dominantly sources CMB temperature fluctuations on small scales, and the hydrogen signal from the patchy reionization. While measuring individual sources is difficult, as they can be confused with other effects, statistical combination of the signals with a common bulk motion, as we discussed here, allows recovering cosmological information at large-scales. The kSZ-reconstructed velocity field can potentially be used to break parameter degeneracies suffered by measurements of the brightness temperature (from 21cm surveys) or the anisotropic optical depth, which can be reconstructed from the measurement of the CMB temperature and polarization, alone. In particular, we see parameters sensitive to time evolution of fluctuations during reionization, i.e. the bubble bias, $b$, the duration of reionziation $\Delta z$ as well as the integrated mean optical depth, $\tau$, see improvement from measurement of the velocity field. {Our improvements compare to similar studies of the kSZ effect from patchy reionization in the literature, such as Ref.~\citep{Alvarez:2020gvl}, for example, where authors find $\sigma(\tau)\!\lesssim\!3\!\times\!10^{-3}$ using kSZ reconstruction internal to CMB.} 
This work illustrates that future CMB and 21cm experiments will provide new ways to test cosmological models and probe inhomogeneities on the very largest scales. 

\noindent\textbf{\textit{Acknowledgements --}\label{sec:acknowledgements}} SCH is funded by Imperial College President's Scholarship, a Visiting Fellowship from the Perimeter Institute for Theoretical Physics, and a postdoctoral fellowship from Imperial College London. We thank Simone Ferraro, Mathew Madhavacheril and James Mertens for their useful comments. We thank Mathew Madhavacheril for providing us with realistic foreground-cleaned CMB noise estimates. Research at Perimeter Institute is supported by the Government of Canada through Industry Canada and by the Province of Ontario through the Ministry of Research \& Innovation. MCJ was supported by the National Science and Engineering Research Council through a Discovery grant.

\bibliography{reio_ksz_main}
\end{document}